\documentclass[preprint,authoryear,12pt]{elsarticle}

\usepackage{amssymb}
\usepackage{amsmath}
\usepackage{comment}

\usepackage[driverfallback=ps2pdf,%
a4paper=true,%
breaklinks=true,%
colorlinks=true,%
pdfauthor={First Author et al.},%
pdftitle={Template for manuscripts in Advances in Space Research}%
]{hyperref}

\journal{Advances in Space Research}

\makeatletter
\def\ps@pprintTitle{%
	\def\@oddhead{\centerline{\scriptsize Accepted for publication in Adv. Space Res. doi: \url{https://doi.org/10.1016/j.asr.2020.08.008}}}
	\let\@evenhead\@empty
	\def\@oddfoot{\centerline{\scriptsize \textcopyright\ 2020. This manuscript is made available under the \href{http://creativecommons.org/licenses/by-nc-nd/4.0/}{CC-BY-NC-ND 4.0 license}}}%
	\let\@evenfoot\@oddfoot}
\makeatother

\begin{document}

\begin{frontmatter}
	
	
	
	\title{DebrisWatch I: A survey of faint geosynchronous debris}
	
	\address[Warwick-astro]{Department of Physics, University of Warwick, Coventry, CV4 7AL (UK)}
	\address[Warwick-hab]{Centre for Exoplanets \& Habitability, University of Warwick, Coventry, CV4 7AL (UK)}
	\address[Dstl-Portsdown]{Defence Science \& Technology Laboratory, Portsdown West, Fareham, PO17 6AD (UK)}
	\address[Dstl-PortonDown]{Defence Science \& Technology Laboratory, Porton Down, Salisbury, SP4 0JQ (UK)}
	\address[SJE]{SJE Space Ltd., Clayhill Road, Burghfield Common, Berkshire, RG7 3HB (UK)}
	\address[QUB]{Astrophysics Research Centre, School of Mathematics and Physics, Queens University, Belfast, BT7 1NN (UK)}
	
	\author[Warwick-astro,Warwick-hab]{James A. Blake\corref{cor}}
	\cortext[cor]{Corresponding author}
	\ead{J.Blake@warwick.ac.uk}
	
	
	\author[Warwick-astro]{Paul Chote}
	\author[Warwick-astro,Warwick-hab]{Don Pollacco}
	
	\author[Dstl-Portsdown]{William Feline}
	\author[Dstl-PortonDown]{Grant Privett}
	
	\author[Dstl-Portsdown]{Andrew Ash}
	\author[SJE]{Stuart Eves}
	\author[Warwick-astro]{Arthur Greenwood}
	\author[Dstl-Portsdown]{Nick Harwood}
	\author[Warwick-astro]{Thomas R. Marsh}
	\author[Warwick-astro,Warwick-hab]{Dimitri Veras}
	\author[QUB]{Christopher Watson}

\begin{abstract}

Recent anomalies exhibited by satellites and rocket bodies have highlighted that a population of faint debris exists at geosynchronous (GEO) altitudes, where there are no natural removal mechanisms.
Despite previous optical surveys probing to around 10--20\,cm in size, regular monitoring of faint sources at GEO is challenging, thus our knowledge remains sparse. 
It is essential that we continue to explore the faint debris population using large telescopes to better understand the risk posed to active GEO satellites. 
To this end, we present photometric results from a survey of the GEO region carried out with the 2.54\,m Isaac Newton Telescope in La Palma, Canary Islands.
We probe to 21$^\text{st}$ visual magnitude (around 10\,cm, assuming Lambertian spheres with an albedo of 0.1), uncovering 129 orbital tracks with GEO-like motion across the eight nights of dark-grey time comprising the survey.
The faint end of our brightness distribution continues to rise until the sensitivity limit of the sensor is reached, suggesting that the modal brightness could be even fainter.
We uncover a number of faint, uncatalogued objects that show photometric signatures of rapid tumbling, many of which straddle the limiting magnitude of our survey over the course of a single exposure, posing a complex issue when estimating object size.
This work presents the first instalment of DebrisWatch, an ongoing collaboration between the University of Warwick and the Defence Science and Technology Laboratory (UK) investigating the faint population of GEO debris.

\end{abstract}


\end{frontmatter}

\parindent=0.5 cm

\section{Introduction}
\label{sec:introduction}

Knowledge of the unique and desirable characteristics of Geosynchronous Earth Orbits (GEOs) predates the dawn of the Space Age. 
Satellites in GEO have an orbital period matching that of the Earth's rotation, meaning that they typically trace a simple analemma (e.g.~an ellipse or a figure of eight) on the sky over the course of a sidereal day (23$^\text{h}$56$^\text{m}$04$^\text{s}$).
In the special case of a geostationary orbit, a station-kept satellite will remain fixed in the observer's sky, a property that has been exploited for communications since the early 1960s.

The GEO region is too high-altitude for atmospheric drag to provide a mechanism for orbital decay, thus there is no natural `sink' for debris residing there.
This is a cause for concern, given that the natural constraints placed on altitude, eccentricity and inclination for the GEO regime already restrict the number of orbital slots.

\begin{figure}[tbp]
	\begin{center}
		\includegraphics*[width=0.75\textwidth]{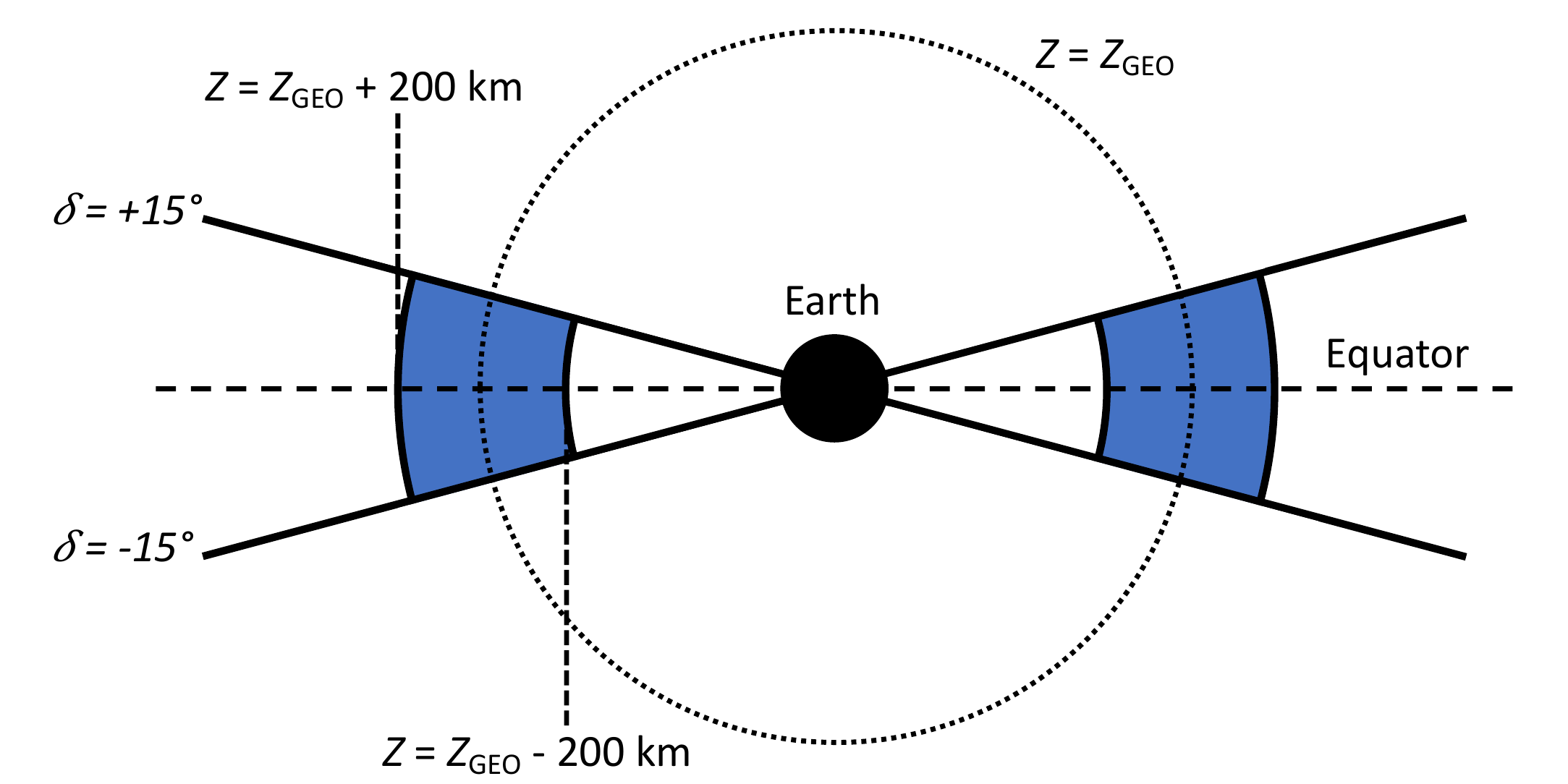}
	\end{center}
	\caption{An illustrative sketch of the GEO Protected Region (in blue), as defined by the IADC Space Debris Mitigation Guidelines~\citep[see][]{iadc2007guidelines}. 
		Extending 200\,km above and below the geostationary altitude $Z_\text{GEO}=35786$\,km, the Region is a segment of a spherical shell spanning $\pm$\,15$^\circ$ in declination relative to the equatorial plane of the Earth.
		Note that the scale of the Protected Region has been exaggerated for clarity.}
	\label{fig:geo-belt-diagram}
\end{figure}

In order to address the problem, guidelines and recommendations~\citep{iadc2007guidelines} have been established over the past two decades to define the GEO Protected Region, depicted in Fig.~\ref{fig:geo-belt-diagram}.
Operators are advised to carry out an end-of-mission (EOM) manoeuvre to a `graveyard' orbit residing outside the Protected Region.
Compliance with the guidelines has improved in recent years, with over 80\,\% of attempted manoeuvres successfully clearing the Protected Region since 2016~\citep{esa2019annual}.

\begin{figure}[tbp]
	\begin{center}
		\includegraphics*[width=\textwidth]{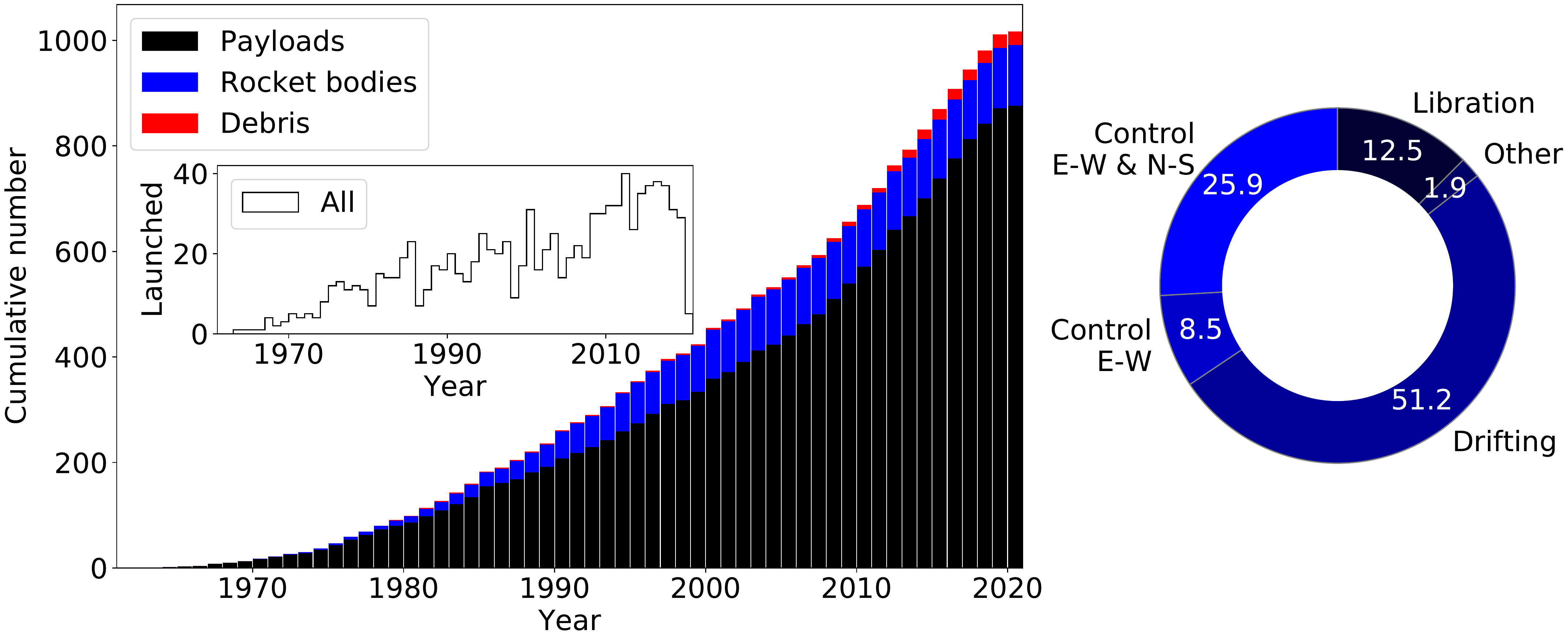}
	\end{center}
	\caption{Left) The cumulative number of tracked objects in GEO, with objects separated into three categories: payloads (black), rocket bodies (blue) and debris (red).
	Debris fragments are binned by the year they were first tracked and catalogued.
	Inset, the number of objects launched to GEO per year.
	Sourced from the publicly available US Strategic Command catalogue, accessed via the geosynchronous sub-catalogue from \href{www.space-track.org}{Space-Track} as of May 2020.
	Right) The orbital status of tracked GEO objects in 2018, as given in~\citet{esa2019annual}.}
	\label{fig:geo}
\end{figure}

In spite of this, it is important to keep in mind the GEO residents that reached EOM prior to the issuance of guidelines, existing in an uncontrolled state ever since.
These are typically in drift orbits or librating about one or both of the geopotential wells that result from the non-spherical shape of the Earth~\citep{mcknight2013new}.
Many of these drift orbits intersect the operational regions of the geostationary belt, posing a direct threat to active satellites.
With the upward trend evident in Fig.~\ref{fig:geo}, it is clear that an imperfect disposal rate will result in fewer orbital slots and increased collision risk in GEO.

Objects in Highly-Eccentric Earth Orbit (HEO) can further add to the risk, with recent observations uncovering a number of fragments penetrating the Protected Region~\citep{schildknecht2019esa}.
Four significant GEO/HEO break-ups have been observed in the past two years alone. 
Collectively, these events produced over 1000 fragments, a few hundred of which cross the GEO Protected Region.
It is also likely that collisions with small debris are responsible for the heavily-publicised anomalies exhibited by the geostationary satellites Intelsat 29e (10/04/2019, NORAD 41308), Telkom 1 (25/08/2017, NORAD 25880) and AMC-9 (17/06/2017, NORAD 27820)~\citep{cunio2017photometric}.

Observations of high-altitude orbits typically employ the use of optical sensors, as their sensitivity drops with the square of range, while that of radar drops more steeply with the fourth power of range.
The publicly available US Strategic Command (USSTRATCOM) catalogue tracks objects in GEO larger than 50-100\,cm, predominantly using a system of 1\,m-class optical telescopes known as Ground-based Electro-Optical Deep Space Surveillance (GEODSS)~\citep{wootton2016enabling}.
Smaller objects are monitored sporadically at best, due to the limited availability of sufficiently sensitive sensors.
This is of particular concern, given a recent study that found relative velocities in GEO can reach up to 4\,kms$^{-1}$, approaching the hypervelocity regime where collisions with cm-sized objects could prove mission-fatal~\citep{oltrogge2018comprehensive}.
As break-ups and anomalies add more small fragments to the GEO environment, it is important that we continue to observe faint objects with large telescopes to better understand their behaviour and the risk they pose to operational satellites.

\begin{table}[tbp]
	\caption{Optical surveys of the GEO region.
		Instrumental fields of view (FOV) are listed as narrow (N) if $\text{FOV}<0.5$\,sq.\,deg, medium (M) if $0.5<\text{FOV}<1$\,sq.\,deg, wide (W) if $1<\text{FOV}<10$\,sq.\,deg and ultra-wide (UW) if $\text{FOV}>10$\,sq.\,deg.
		Survey depths are denoted $M_{X}$ for a given photometric band $X$, and $M$ for cases where the band is not specified or absolute magnitudes have been quoted~\citep[see][]{africano2005phase}.}
	\begin{center}
		\begin{tabular}{lrrrr}
			\hline
			Survey & Instr. & Instr. & Survey & Reference \\
			& size [m] & FOV & depth & \\
			\hline
			NASA CDT & 0.32 & W & $M\sim16$ & \citet{barker2005cdt} \\
			MODEST & 0.61 & W & $M_R\sim17$ & \citet{seitzer2004modest} \\
			TAROT & 0.18--0.25 & W--UW & $M_R\sim15$ & \citet{alby2004status} \\
			ESA-AIUB & 1.00 & M & $M_V\sim20$ & \citet{schildknecht2007optical} \\
			ISON & 0.22--0.70 & N--UW & $M\sim18$ & \citet{molotov2008international} \\
			ISON (faint) & 1.00--2.60 & N & $M\sim20$ & \citet{molotov2009faint} \\
			Pan-STARRS & 1.80 & UW & $M_V\sim21$ & \citet{bolden2011panstarrs} \\
			Magellan & 6.50 & N & $M_R\sim19$ & \citet{seitzer2016the} \\
			FocusGEO & (3$\times$)0.18 & UW & $M\sim15$ & \citet{luo2019focusgeo} \\
			\hline
		\end{tabular}
	\end{center}
	\label{tab:surveys}
\end{table}

We provide an overview of past/ongoing GEO surveys in Table~\ref{tab:surveys}.
The majority have utilised optical telescopes with diameters of 1\,m or less, with sensitivity limits in the range 15$^\text{th}$--20$^\text{th}$\,Magnitude, corresponding to objects larger than $\sim$15\,cm in diameter (depending on the viewing geometry, assumed shape and reflectivity).
A small number of surveys have uncovered fainter objects with telescopes larger than 1\,m.
For example, the 6.5\,m Magellan telescope has been used for a small number of GEO spot surveys, targeting known fragmentation events~\citep{seitzer2016the}.
These deeper observations, alongside those conducted by the Panoramic Survey Telescope and Rapid Response System (Pan-STARRS) 1.8\,m~\citep{bolden2011panstarrs} and large-aperture telescopes of the International Scientific Optical Network (ISON)~\citep{molotov2009faint}, have found that many faint detections show photometric signatures of tumbling.

In this paper, we present photometric results from a survey of the GEO region undertaken with the 2.54\,m Isaac Newton Telescope in La Palma, Canary Islands. 
The survey was carried out as part of DebrisWatch, an ongoing collaboration between the University of Warwick and the Defence Science and Technology Laboratory (UK) investigating the faint debris population at GEO.
We outline our observational strategy in Section~\ref{sec:observational-strategy}, optimised for finding objects in GEO.
Section~\ref{sec:analysis-pipeline} provides an overview of our data analysis pipeline, which performs reduction tasks, object detection and light curve extraction.
In Section~\ref{sec:results-discussion}, we consider the population sampled by our observations and present light curves for objects of interest, before discussing our findings and future plans.

\section{Observational strategy}
\label{sec:observational-strategy}

\begin{table}[tbp]
	\caption{Logistical details for the observation run. 
	In total, 552 separate pointings of the telescope in hour angle and declination were achieved in the 58 hours of survey time. 
	Approximately half the night of 5$^\text{th}$ September was lost due to weather and technical issues. 
	The remaining time was dedicated to targeted observations that are outside the scope of this work.}
	\begin{center}
		\begin{tabular}{lrr}
			\hline
			Night & Survey time & Telescope \\
			  & [hrs] & pointings \\
			\hline
			02/09/2018 & 8.5 & 65 \\
			03/09/2018 & 7.7 & 76 \\
			04/09/2018 & 6.4 & 71 \\
			05/09/2018 & 4.5 & 15 \\
			06/09/2018 & 7.0 & 77 \\
			07/09/2018 & 6.0 & 63 \\
			08/09/2018 & 8.5 & 86 \\
			09/09/2018 & 9.4 & 99 \\
			\hline
			& 58.0 & 552 \\
			\hline
		\end{tabular}
	\end{center}
	\label{tab:logistics}
\end{table}

We used eight nights of dark-grey time on the 2.54\,m Isaac Newton Telescope (INT) to conduct an untargeted survey of the GEO region visible from the Roque de los Muchachos Observatory in La Palma, Canary Islands.
Logistical details for the survey are provided in Table~\ref{tab:logistics}.

Observations were made using the prime focus Wide Field Camera (WFC), consisting of four thinned 2k\,$\times$\,4k charge-coupled device (CCD) chips, which combine to image over a 33$'$ field of view.
One of the CCD chips was rendered unusable due to an issue with the readout electronics.
We discard this chip for the following photometric analyses, reducing our effective field of view to 22$'$\,$\times$\,33$'$.
Two-by-two binning was applied, resulting in a resolution of 0.66$''$pixel$^{-1}$.
The observations were taken using a Harris V filter with a central wavelength of 5425\,{\AA}, a full width at half maximum of 975\,{\AA} and a peak throughput of 88\,\%.

Steps were taken to optimize our observations for finding objects in GEO.
The telescope was operated at a fixed hour angle and declination, ensuring that photons from GEO candidates would integrate across fewer pixels to improve the signal-to-noise ratio.
In this observing mode, GEO objects manifest as point sources or short trails in the resulting image, while background stars appear as longer trails, streaking across at the sidereal rate.
We chose an exposure time of 10\,s to provide a balance between streak coverage and duty cycle.
Observations were taken by selecting a nominal field with a specific right ascension and declination, corresponding to a fixed solar phase angle (observatory-target-Sun), which was minimised whilst remaining outside of the Earth's shadow.
This allowed for the detection of fainter objects by maximising their apparent brightness.
The selected field would then be used to generate the telescope pointings for the given night, scanning a strip of fixed declination with each pointing fixed at a separate hour angle.

\begin{figure}[tbp]
	\begin{center}
		\includegraphics*[width=\textwidth]{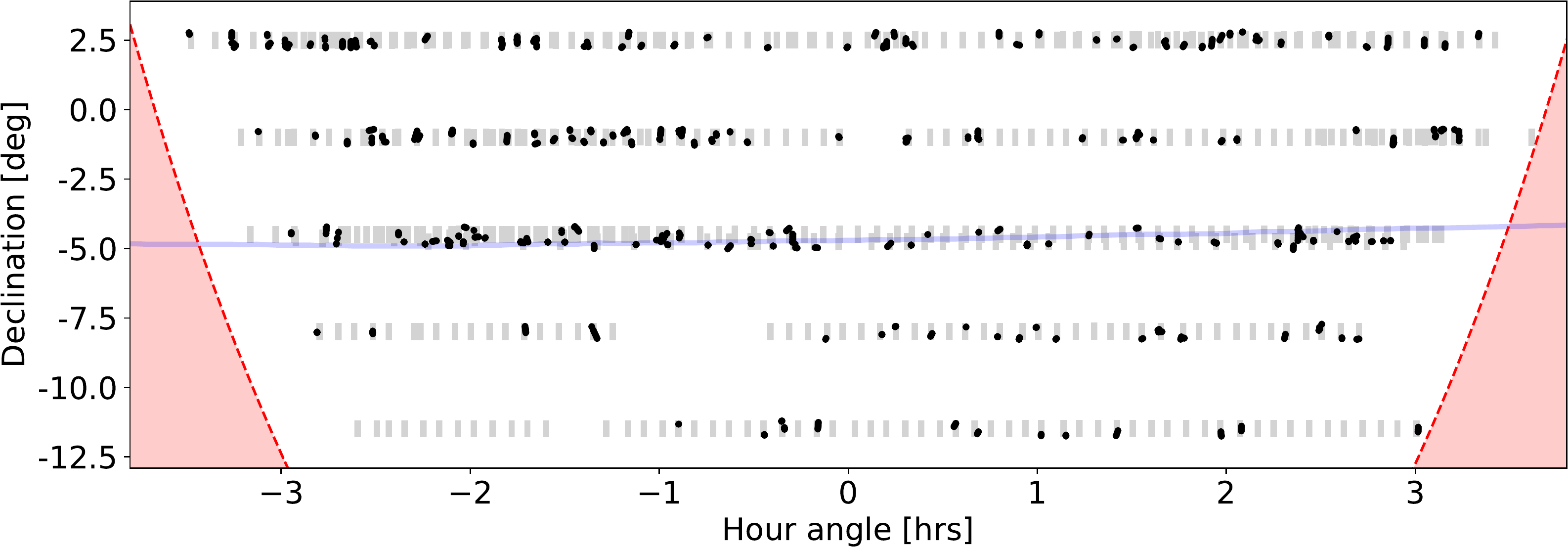}
	\end{center}
	\caption{Telescope pointings for the survey.
	Imaged fields are given by the grey boxes, while detections are overlaid as black dots.
	The approximate declination of the geostationary belt as visible from the vantage point of the INT is indicated by the blue line.
	Shaded red regions mark the altitude limits that constrained the accessible range in hour angle for a given declination strip.}
	\label{fig:target-fields}
\end{figure}

We provide a map of telescope pointings in hour angle and declination for the survey in Figure~\ref{fig:target-fields}.
The INT telescope control system disables several important instrument features upon issuance of a telescope stop command.
We instead applied a differential tracking offset upon reaching the chosen field, in order to counter the sidereal rate and freeze the hour angle for the duration of the given pointing.
Each telescope pointing was observed for roughly four minutes, comprising seven 10\,s exposures with a 25\,s readout time per exposure.
Multiple exposures were taken at each pointing to allow for correlation of detections across frames.
After each set of exposures, the telescope pointing was updated to retrieve the chosen field and the above procedure was repeated.
Survey operations began when the target field exceeded 30$^\circ$ elevation in the east and continued until it set below 30$^\circ$ elevation in the west.
Most aspects of the observing procedure were automated using a script, however limitations in the INT control system meant that operator input was required for each new pointing.

The observation script also sent commands to a second telescope on-site, a 36\,cm astrograph assembled from commercial-off-the-shelf (COTS) equipment, featuring a much larger 3.6$^\circ$\,$\times$\,2.7$^\circ$ field of view.
The astrograph remained slaved to the INT for the duration of the observation campaign. 
The additional dataset from this instrument will form the basis of a future DebrisWatch study that will test the capabilities of COTS hardware against those of large telescopes when tasked with detecting faint objects in GEO.

\section{Analysis pipeline}
\label{sec:analysis-pipeline}

\begin{figure}[tbp]
	\begin{center}
		\includegraphics*[width=\textwidth]{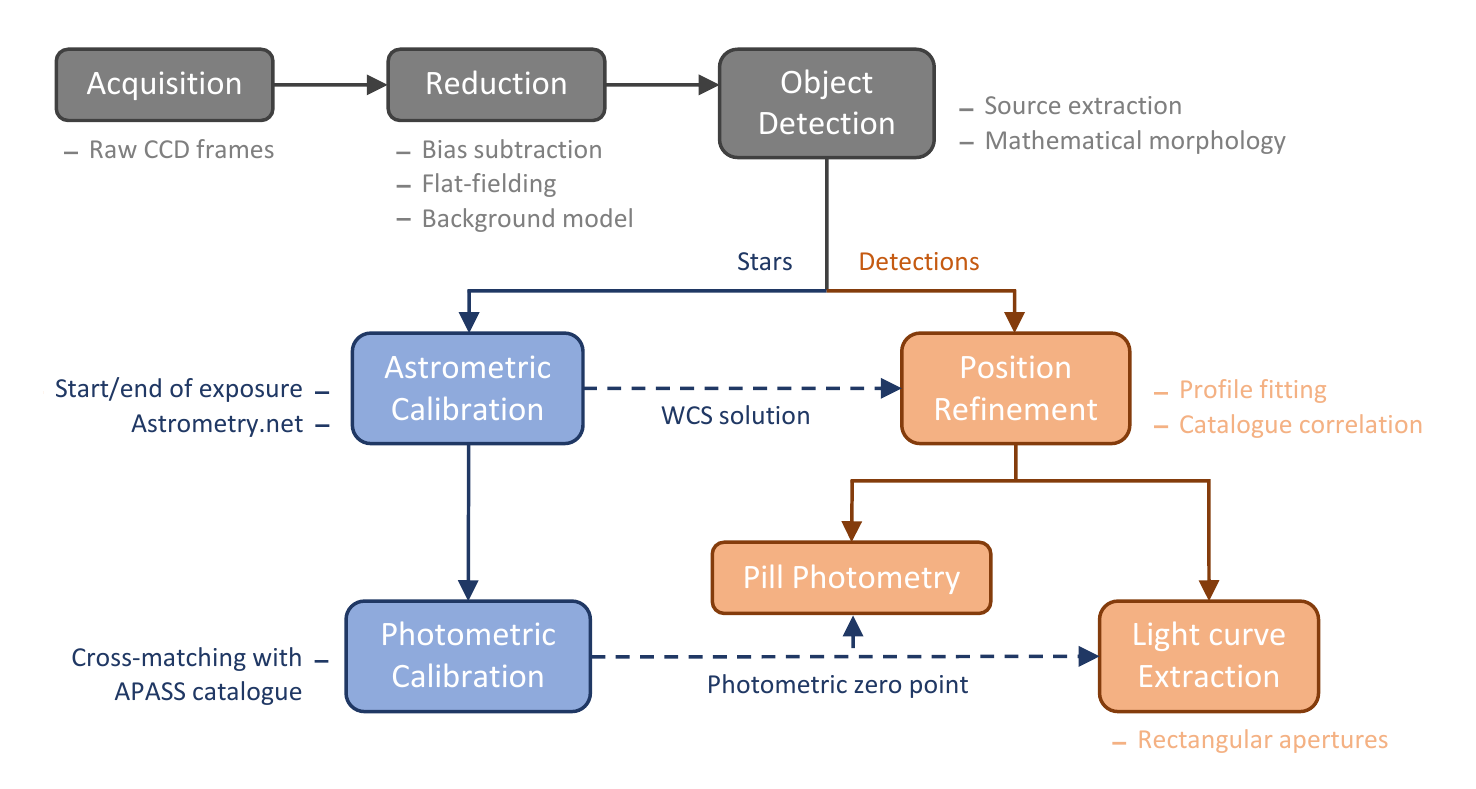}
	\end{center}
	\caption{Diagram outlining the analysis pipeline described in Section~\ref{sec:analysis-pipeline}.}
	\label{fig:flowchart}
\end{figure}

The survey data were processed using a custom analysis pipeline, which is outlined in Fig.~\ref{fig:flowchart}.
Written in Python 3, the pipeline takes inspiration from a number of algorithms developed previously to find artificial objects in astronomical images~\citep{laasbourez2009tarot,levesque2009automated,privett2017autonomous}.

\subsection{Reduction}
\label{sec:reduction}

Standard bias and flat-field calibrations are applied using calibration frames acquired at the beginning of each night.
A bad pixel mask was created from the flat-field observations and defective pixels in the science frames were replaced with a sigma-clipped median of the surrounding pixel values.
We use \texttt{SEP} (Source Extractor in Python) to subtract a model of the spatially-varying sky background from the calibrated frame~\citep{barbary2016sep,bertin1996sextractor}.

We then utilise the extraction capabilities of \texttt{SEP} to find stars in the image, exploiting their common morphologies and orientations.
The centroids and start/end points of the star trails are fed to \textit{Astrometry.net}, which pattern-matches subset quadrilaterals of stars against sky catalogues to determine accurate World Coordinate System (WCS) solutions for astronomical images~\citep{lang2010astrometry}. 
Following this, it is simple to convert between pixel and sky coordinates using \texttt{astropy} WCS routines~\citep{pricewhelan2018astropy,robitaille2013astropy}.

Using this astrometric solution, we perform a photometric calibration by cross-matching the star trails with the American Association of Variable Star Observers (AAVSO) All-Sky Photometric Survey (APASS) catalogue~\citep{henden2016apass}.
The photometric zero point for the frame is found by comparing the standard magnitudes quoted in the APASS catalogue against their instrumental counterparts, derived by summing rectangular apertures placed over the star trails.

\subsection{Object detection}
\label{sec:object-detection}

\begin{figure}[tbp]
	\begin{center}
		\includegraphics*[width=\textwidth]{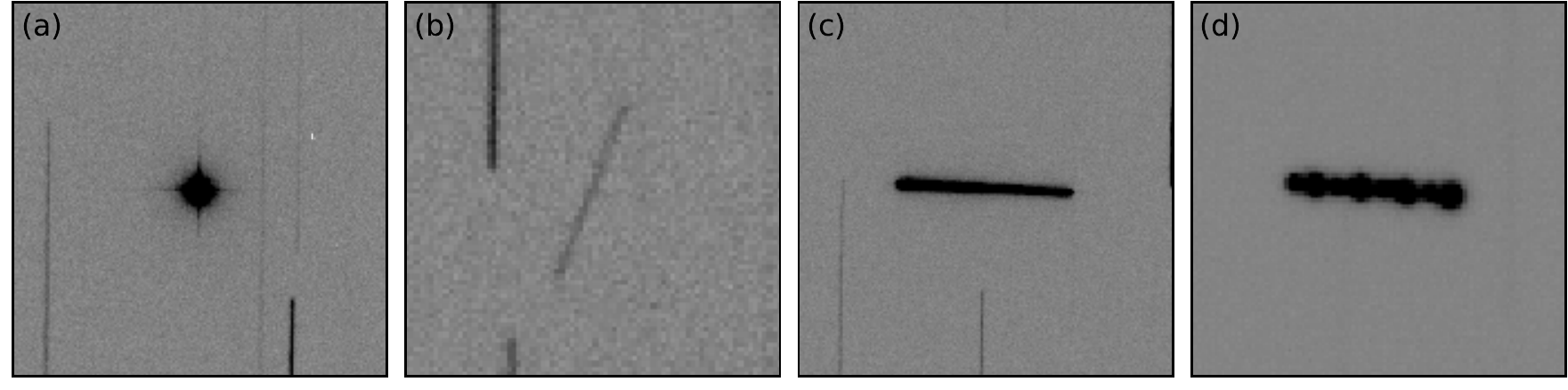}
	\end{center}
	\caption{Examples of object morphologies.
	With the telescope viewing direction fixed relative to the Earth, geostationary satellites appear as point-like features in the acquired CCD frames, as in (a).
	Objects in GEO that are moving relative to geostationary will manifest as trails.
	An example of a faint trail with uniform brightness is given in (b).
	Other trails exhibit brightness variation, over timescales longer (c) and shorter (d) than the exposure time.
	Vertical lines in the background are stars streaking across the images.}
	\label{fig:morphology}
\end{figure}

Many GEO residents are moving relative to the geostationary tracking rate and so are not fixed in the topocentric coordinate frame.
The reflected light from these objects will spread over a trail of pixels mapped out by the angular path traversed during the exposure.
Additional structure along the trails (e.g. glints, flares, gentle oscillations) can result from changes in the reflected light received from the object along the observer's line of sight. 
As a result, objects of interest exhibit a wide range of morphologies and orientations, examples of which can be seen in Fig.~\ref{fig:morphology}.

We remove the background star trails using mathematical morphology, a technique for examining geometrical structures within images~\citep{breen2000mathematical,matheron2002mathematical}.
As in~\citet{laasbourez2009tarot}, we probe each image $f(x)$ with a structuring element $B$ using the Spread TopHat transformation $\eta$,
\begin{equation}
\eta^B(f(x))=f(x)-O^B(C^B(f(x))).
\end{equation}
The opening $O$ and closing $C$ operations act to remove small peaks and dark regions, respectively.
When combined to form the Spread TopHat, the effect is to remove features that contain the structuring element, whilst limiting remnant noise in the resulting image.
We carry out the transformation using the \texttt{scipy} morphology routines~\citep{jones2001scipy}.
Rectangular structuring elements are used to emulate the star streaks in our images, with dimensions 1\,$\times$\,$\frac{1}{2}l_\text{ST}$\,px for the opening and 1\,$\times$\,$\frac{1}{6}l_\text{ST}$\,px for the closing, given an expected star trail length $l_\text{ST}$.
Candidate GEO objects are retained as they do not contain either of the structuring elements.


Additional checks are required to separate the objects of interest from remnant `distractors' that survive the transformation.
After running the \texttt{ccdproc} lacosmic routine~\citep{craig2015ccdproc,vandokkum2001cosmic} to remove cosmic rays from the transformed image, we apply a 3$\sigma$ threshold cut to filter out the majority of spurious detections, where $\sigma$ is the global background root mean square.
The remaining false positives are typically edges of star trails that are easily flagged given our knowledge of the trail positions.

\subsection{Position refinement}
\label{sec:position-refinement}

In the case of trailed detections, it is necessary to accurately determine the start and end points, as we know these will correspond to the angular positions of the object at the start and end of exposure, respectively.
We refine the initial estimate from a \texttt{SEP} extraction by fitting the intensity profiles along and across the trail.
We use a Gaussian fit for the across-trail intensity profiles, while a good approximation of the along-trail profile is given by the `Tepui' function,
\begin{equation}
I(x)=A\left[\arctan(b_1(x-c-x_0))-\arctan(b_2(x+c-x_0))\right],
\end{equation}
where $A$ is the normalised amplitude, $b_1$ and $b_2$ are related to the profile tilt, $c$ gives the half-width and $x_0$ is a translational offset. 
Several studies have made use of the Tepui function when fitting streaks in astronomical images~\citep[see e.g.][]{lacruz2018astrometric,montojo2011astrometric,park2016development}.

Using this refinement procedure, we obtain typical uncertainties of 1-2$''$, corresponding to 200-400\,m at GEO.
Within the scope of our photometric study, this level of uncertainty was deemed acceptable. 
We use the refined estimate of the orientation to predict where the object will appear in subsequent frames within a given pointing, correlating trails belonging to the same orbital track.
In the photometric analyses that follow, we only consider objects that appear in two or more frames.

The refined orientation allows for more accurate placement of a \texttt{TRIPPy} pill aperture~\citep{fraser2016trippy}, the sum of which provides a measure of the total flux integrated over the course of the exposure.
Trail morphologies are well-approximated by pill shapes, so the contribution of background noise to the aperture sum is minimised.
Uncertainties in the measured magnitudes consist of two parts: the first is a systematic uncertainty from the zero point measurement, which is based on the background stars and is typically $\sim$0.05\,mag for a given frame, while the second is the photometric uncertainty from the aperture sum, which is typically $\sim$0.001\,mag for bright objects ($V\sim12$) and $\sim$0.05\,mag for faint objects ($V\sim18$) in a 10\,s exposure.
We note that intrinsic brightness variability can cause much larger scatter in short-timescale measurements for specific objects, as will be illustrated in Section~\ref{sec:photometric-light-curves}, where we provide examples of light curves extracted from our detections.

\subsection{Light curve extraction}
\label{sec:light-curve-extraction}

In the final stage of the pipeline, we extract light curves from our trailed detections.
Rectangular apertures are placed along the trail, each covering a discrete pixel in width to avoid correlated noise injection.
We assume constant rates of change in angular position throughout the exposure.
Background contamination (e.g. blending with star streaks) is corrected by placing equivalent apertures in a reference frame containing the same field.
We perform an initial image alignment using the astrometric solutions for the frames, then account for remnant offsets using the \texttt{DONUTS} alignment algorithm~\citep{mccormac2013donuts}.

\section{Results and discussion}
\label{sec:results-discussion}

\subsection{Sampled population}
\label{sec:sampled-population}

\begin{figure}[tbp]
	\begin{center}
		\includegraphics*[width=\textwidth]{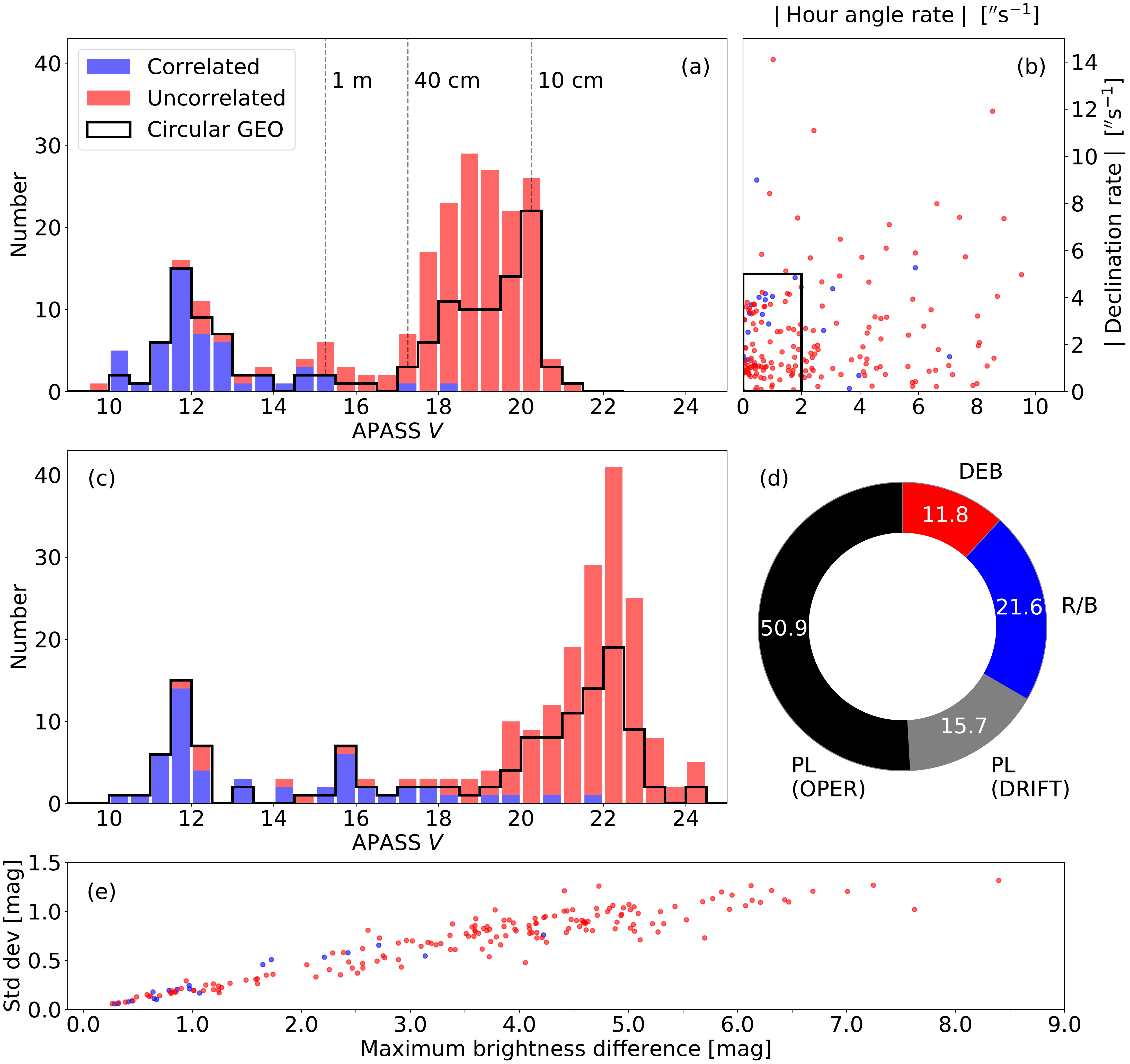}
	\end{center}
	\caption{(a) Brightness histogram for the detected population, as previously presented in~\citet{blake2019optical}.
	Tracks that correlate with the publicly available USSTRATCOM catalogue are shown in blue, while those that fail to correlate are in red.
	The black line gives our sub-sample of tracks which lie within the rate cut limits shown in (b).
	Labelled size estimates assume that the objects are Lambertian spheres with an albedo $A=0.1$.
	(b) Rates in hour angle and declination for the tracks detected.
	The rate cuts applied in order to obtain the circular-GEO sub-sample are indicated by the black box.
	(c) Brightness histogram for the detected population, normalised by trail length.
	This normalisation determines the brightness of a geostationary source with the same peak flux for an equivalent integration time.
	(d) Object types for correlated detections in the overall sample, categorised as operational or drifting payloads (PL), rocket bodies (R/B) or debris (DEB).
	(e) Light curve statistics for the overall sample.
	For each detected object, the difference between the maximum and minimum brightness is plotted against the standard deviation.}
	\label{fig:brightness-rates}
\end{figure} 

A total of 226 orbital tracks spanning two or more exposures within a given pointing were detected.
The brightness distribution for these detections is presented in panel (a) of Fig.~\ref{fig:brightness-rates}.
We limit our attention to tracks that are consistent with circular orbits in the GEO regime, using the cuts defined in~\citet{seitzer2011faint}:
\begin{equation}
\lvert\text{Hour Angle rate}\rvert<\;2''\text{s}^{-1}\; \text{and}\; \lvert\text{Declination rate}\rvert<\;5''\text{s}^{-1}.
\end{equation}
Objects with rates exceeding these limits likely reside in geosynchronous transfer orbits (GTOs) which are elliptical orbits with apogees in the GEO region.
The resulting subset of circular-GEO detections is represented by the black lines in Fig.~\ref{fig:brightness-rates}.

We correlate our detections against the publicly available USSTRATCOM catalogue, finding that 85\,\% of tracks with $V<15$ successfully correlate, while only 1\,\% of fainter detections match a known object.
This is consistent with the $\sim$1\,m cut-off for the GEODSS network.

The rate cuts reduce our sample size to 129 circular-GEO tracks, giving a detection rate of $\sim$11\,hour$^{-1}$deg$^{-2}$ for the survey.
A similar detection rate was observed by the Magellan surveys in Chile~\citep{seitzer2011faint}, within sight of the geopotential well at longitude 105$^\circ$\,W. 
Risk assessments have found that collision probabilities increase by a factor of seven in the vicinity of the potential wells~\citep{mcknight2013new}, owing to the relatively high density of trapped objects in libration orbits.
La Palma ($\sim$18$^\circ$\,W) sits almost directly between the two wells, thus we would expect to have a lower detection rate.
However, the limited time available on large telescopes means that both surveys suffer from small number statistics, making it difficult to draw conclusions regarding detection rates at this early stage.

We observe a bimodal brightness distribution, consistent with the findings of previous GEO surveys.
The bright end of the sample peaks at $V\sim12$, in accordance with the population uncovered by the European Space Agency (ESA) 1\,m Optical Ground Station (OGS) observation campaigns in Tenerife~\citep{schildknecht2004optical}.
This is to be expected given that the majority of bright, correlated objects are geostationary and the two instruments sample the same section of the GEO belt.
For reference, we classify our correlated detections according to object type in panel (d) of Fig.~\ref{fig:brightness-rates}.
We see a steep rise in the number of objects detected as we look fainter than $V\sim17$.
The overall distribution appears to plateau between $V\sim18$ and our sensitivity limit at $V\sim21$.
Our circular-GEO sample continues to rise as we reach the sensitivity limit, suggesting that the modal brightness may be fainter still.

Assuming the objects are Lambertian spheres with albedo $A=0.1$, we probe to sizes $d<10$\,cm~\citep{africano2005phase}.
These assumptions are nevertheless very uncertain, as we lack \textit{a priori} knowledge for any object that fails to correlate with the catalogue.
Furthermore, the brightness of a given object is not always constant over the course of an observation.
Indeed, from panel (e) of Fig.~\ref{fig:brightness-rates}, we see that over 45\,\% of uncorrelated tracks in the overall sample with successfully extracted light curves vary in brightness by more than 4\,mag across the observation window. 
In some cases, such brightness variation may manifest as sharp flares or glints, while other objects may exhibit smooth oscillations between successive maxima and minima. 
Photometric behaviour of this kind renders any generalisation regarding the albedo redundant.
We find that uncorrelated detections appear to show a greater extent of brightness variation relative to their correlated counterparts within the sampled population.

In addition, the apparent sensitivity limit in panel (a) of Fig.~\ref{fig:brightness-rates} is not truly representative of the detection capability of the sensor, as intrinsic brightness will not be the only factor influencing this.
As revealed by our rate cuts, many objects have non-zero rates of change in angular position, placing a limit on the amount of time they will spend contributing flux to a given set of pixels and therefore reducing the peak surface brightness.
To highlight this effect, we normalise the total flux integrated for each of our detections by a factor $x/l$, where $x$ is characteristic of the point spread function (PSF) of the optical system and $l$ is the extent of the angular path mapped by the object over the course of the 10\,s exposure. 
This normalisation gives the brightness of a point-like detection that would possess an equivalent peak flux for the same integration time, resulting in the updated brightness histogram in panel (c) of Fig.~\ref{fig:brightness-rates}.
The faint end of the circular-GEO distribution now peaks at $V\sim22$, before dropping off as we reach our sensitivity limit for `stationary' objects, implying that the modal brightness in this normalised regime could once again be even fainter.
With the INT, we achieve $x=3.3$\,px, meaning that an object moving at the maximum angular rate allowed by our cuts would take 0.4\,s to cross each pixel.
Exposing for longer than this time will weaken the ability of the pipeline to detect such an object, due to added noise from the sky background.

\begin{figure}[tbp]
	\begin{center}
		\includegraphics*[width=\textwidth]{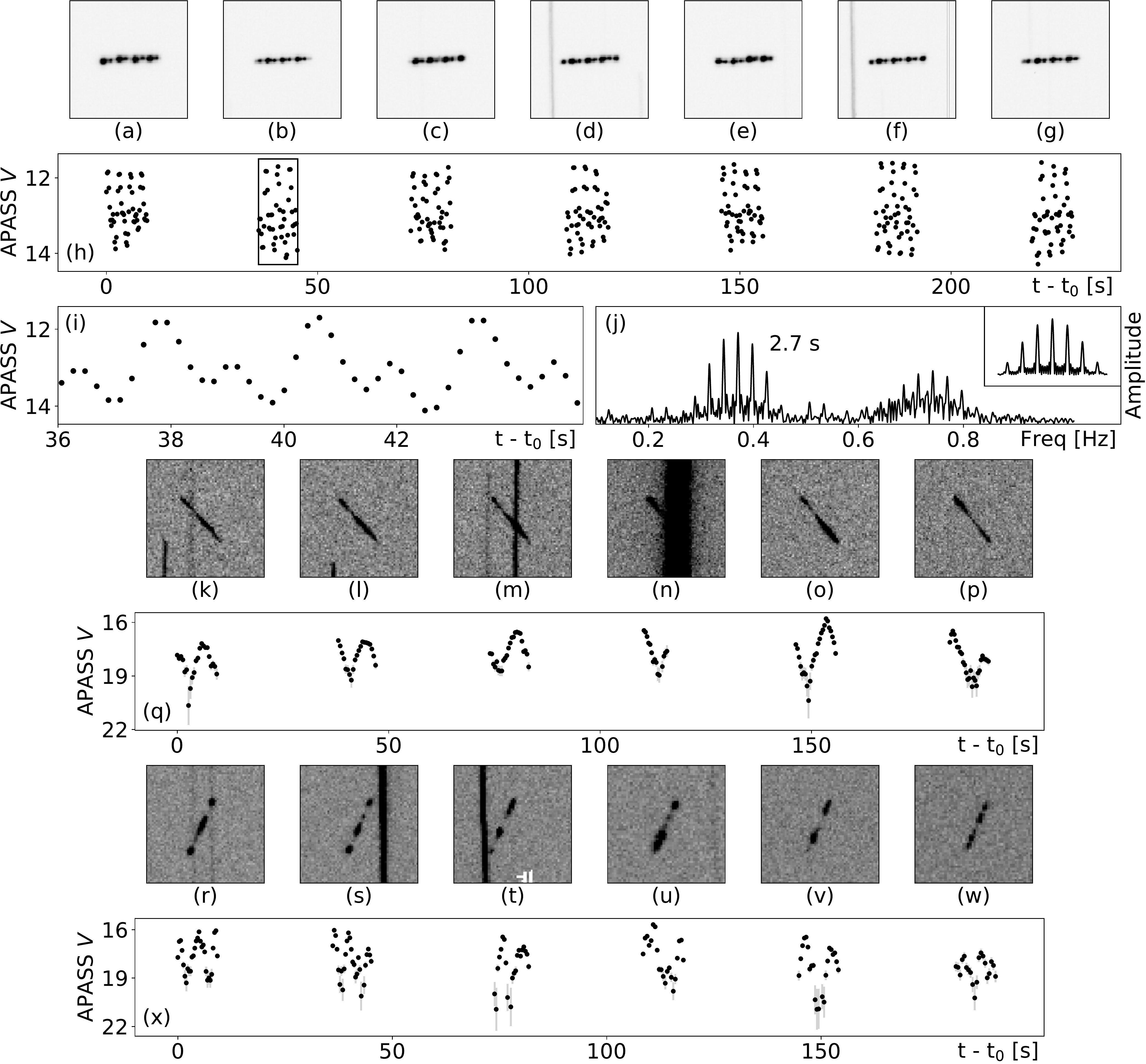}
	\end{center}
	\caption{The top three rows present light curve analysis for a track correlated with the defunct satellite SBS-3 (NORAD 13651).
	Successive 10\,s exposures are shown in (a)--(g), centered on the detected trails.
	The corresponding light curve is provided in (h), extracted using the analysis pipeline outlined in Section~\ref{sec:analysis-pipeline}, while a zoom-in of the boxed region is given in (i).
	A 2.7\,s period is uncovered by the Fourier amplitude spectrum in (j).
	The Fourier window function, displayed inset, illustrates the effect of the readout-induced gaps in the light curve.
	The remaining rows present light curves for two uncorrelated tracks that exhibit significant brightness variation.
	Successive 10\,s exposures are shown in (k)--(p) for the first object, while (q) gives the extracted light curve.
	Note that the 10\,s exposure images provided in (r)--(w) for the second track are reflected in the horizontal direction, aligning each trail with its corresponding profile in (x).
	The three examples shown are as presented previously in~\citet{blake2019optical}.}
	\label{fig:lightcurves-original}
\end{figure}

\subsection{Photometric light curves}
\label{sec:photometric-light-curves}

The reflected light from an orbiting body contains information about its shape and attitude, but is also affected by the sensor characteristics, atmospheric interference and the viewing geometry at the time of the observation.
Disentangling these components is a difficult task and light curve characterisation remains an active area of research~\citep[see e.g.][]{albuja2018yorp,cognion2014rotation,fan2019inversion,hinks2016angular,papushev2009investigations}.
Thus far, studies have focused on modelling the photometric signatures of large satellites by virtue of the relative ease in obtaining a useful dataset.
However, understanding the attitude of faint objects will be a pivotal factor in predicting the long-term evolution of the GEO debris environment.

An example of a light curve extracted for a catalogued object can be found in panel (h) of Fig.~\ref{fig:lightcurves-original}.
The corresponding orbital track correlates with SBS-3 (NORAD 13651), a decommissioned communications satellite that was moved to a graveyard orbit in 1995.
Built on the Hughes HS-376 bus, the satellite consists of a cylindrical body with concentric solar panels and extended antennas.
The satellite was spin-stabilised during its active lifetime, maintaining attitude by spinning a section of the platform at 50 rpm (0.83\,Hz; 1.2\,s period).
The communications payload remained despun, ensuring steady pointing of the antennas and transponders.
A periodic pattern can be seen in the light curve, indicating that the satellite is likely tumbling.
Fourier analysis of the signal uncovers a 2.7\,s period for the repeated pattern, though this could be a harmonic of the true tumbling rate given the geometric symmetry of the bus.

In panels (q) and (x) of Fig.~\ref{fig:lightcurves-original}, we show two examples of light curves extracted for uncorrelated objects belonging to the faint end of our sampled population.
Both tracks straddle the sensitivity limit of our observations, exhibiting significant brightness variation across the observation window.
The first object oscillates in brightness with a period similar to the exposure time, peaking at $V\sim16$ and otherwise fading into the background noise level.
With such large variation in brightness, it is likely that the object is a small piece of highly-reflective material tumbling in and out of our line of sight.
Additional structure can be seen in the second light curve, possibly due to an asymmetry in the shape, or more complex tumbling dynamics.

\begin{figure}[tbp]
	\begin{center}
		\includegraphics*[width=\textwidth]{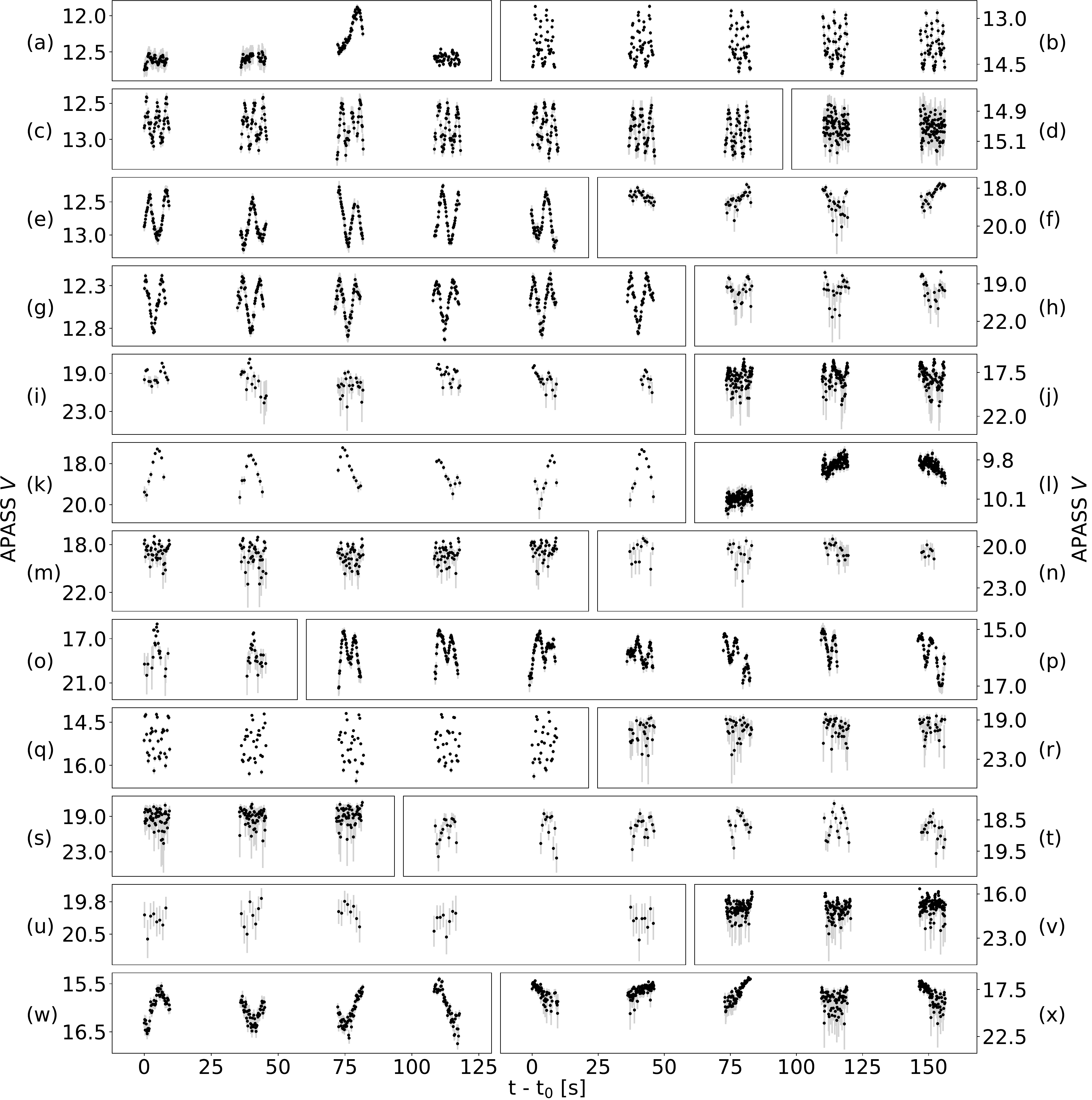}
	\end{center}
	\caption{A montage of light curves for orbital tracks comprising the sampled population, extracted using the analysis pipeline presented in Section~\ref{sec:analysis-pipeline}.}
	\label{fig:lightcurve-montage}
\end{figure}

We provide a montage of further light curve examples in Fig.~\ref{fig:lightcurve-montage}.
Light curve (a) corresponds to a bright orbital track that correlates with Raduga 13 (NORAD 14307), a former Soviet communications satellite that was launched in 1983 and now resides in a drift orbit.
The satellite is based on the KAUR-3 bus, a three-axis stabilised `box-wing' model with solar panels extending from both sides of the main body.
We see a relatively flat light curve at $V\sim12.5$ across all but one exposure, which captures a clear glint where a highly reflective component enters the line of sight.
The light curve in panel (b) is that of a track correlated with Intelsat 4A-F3 (NORAD 10557), a retired communications satellite that launched in 1978.
Based on the Hughes HS-353 platform, the lightcurve unsurprisingly exhibits similar photometric signatures to those of SBS-3 presented in Fig.~\ref{fig:lightcurves-original}.

Panels (c), (e) and (g) of Fig.~\ref{fig:lightcurve-montage} give the light curves for three SL-12 rocket bodies (NORAD 16797, 15581 and 23883, respectively).
Fourier analysis of light curve (c) uncovers a period of 3.4\,s; the SL-12 appears to exhibit higher-frequency brightness variations than expected from previous studies of such rocket bodies~\citep[see e.g.][]{cardona2016bvri}, though aliasing effects could be at play as a result of the object's geometric symmetry.
The $\sim5$~s period signals obtained for the other two SL-12 light curves are in better agreement with the findings of the cited study.

The remaining light curves in Fig.~\ref{fig:lightcurve-montage} correspond to orbital tracks that fail to correlate with catalogued objects.
Light curves (f), (k), (l), (u), (w) and (x) all appear to be oscillating in brightness with a period exceeding the exposure time of 10\,s.
In these cases, it would be necessary to follow-up with targeted observations of the object, preferably using an instrument with reduced dead time, in order to gain confidence in the true profile.
We also find a number of uncorrelated objects that show structure in their light curves on a timescale shorter than the exposure time; this is the case for light curves (p), (q) and (t).

An interesting group of detections uncovered by the survey are only detectable as a result of sharp glints that can occur several times per exposure.
Examples of this behaviour can be found in panels (h), (i), (j), (m), (o), (s) and (v).
The extent of the brightness increase during a glint varies significantly case-by-case, with some objects climbing in excess of 5\,mag above the sensitivity limit, while others struggle to breach it.
Finally, light curves (d), (n) and (r) show little variation in brightness within the window of observation.
There are several explanations as to why this may be the case.
The corresponding object could be uniformly reflective across its surface, or oriented in such a way that higher-reflectivity components were hidden from our line of sight for the duration of the pointing.
Alternatively, the object may be stable in its motion (unlikely for the very faint examples) or tumbling faster than the sampling rate of our observations, such that photometric signatures are unresolved. 
Noisy scatter could be due to small sub-structures upon the object's surface, although atmospheric fluctuations will also contribute to noise in all of our light curves.


\section{Conclusion}
\label{sec:conclusion}

We conducted an optical survey of the GEO region with eight nights of dark-grey time on the 2.54\,m Isaac Newton Telescope (INT) in La Palma, Canary Islands.
Using an optimised observational strategy (see Section~\ref{sec:observational-strategy}) and a custom analysis pipeline (see Section~\ref{sec:analysis-pipeline}), we found:
\begin{itemize}
	\item a total of 226 orbital tracks, 129 of which exhibit rates of change in angular position consistent with circular orbits in the GEO regime;
	\item a detection rate of $\sim$11\,hour$^{-1}$deg$^{-2}$ for circular-GEO objects, similar to rates observed by the Magellan spot surveys of GEO;
	\item a bimodal brightness distribution, with the bright end centered around $V\sim12$ and the faint end still rising at our sensitivity limit of $V\sim21$, suggesting the modal brightness may be fainter still;
	\item over 80\,\% of tracks with $V<15$ correlated with objects in the publicly available USSTRATCOM catalogue, while the vast majority of fainter tracks failed to correlate;
	\item many faint, uncorrelated objects show optical signatures of tumbling, causing some to straddle the detection limit of our observations within a single exposure.
\end{itemize}

The GEO region is an important commodity with a limited number of orbital slots.
Free slots are set to become increasingly scarce with an imperfect disposal rate and an increase in orbital break-ups and anomalies in recent years.
The latter have injected over a thousand new fragments into high-altitude orbits since 2018, with a few hundred intersecting the GEO Protected Region.
The majority of these fragments are too faint to be tracked and made publicly available via the USSTRATCOM catalogue, with its size cut-off of $\sim$50--100\,cm at GEO.
It is therefore essential that we probe the faint end of the debris population to gain a better understanding of the GEO environment both in the short- and long-term.

The presented survey was carried out as part of DebrisWatch, an ongoing collaboration between the University of Warwick and the Defence Science and Technology Laboratory (UK) investigating the faint population of GEO debris.
For the duration of the observation campaign, a 36\,cm astrograph was slaved to the INT, covering the same regions of sky with a larger field of view. 
Analysis of this rich dataset is ongoing and will form the basis of future DebrisWatch instalments.

\section*{Acknowledgements}
\label{sec:acknowledgements}
JAB gratefully acknowledges support from the STFC (grant ST/R505195/1). PC acknowledges support by the STFC via an IPS Fellowship (grant ST/R005125/1).
DV is also supported by the STFC via an Ernest Rutherford Fellowship (grant ST/P003850/1).
DP acknowledges the Royal Society for support.
TRM acknowledges support from the STFC (grant ST/P000495/1).
This paper makes use of data from the Isaac Newton Telescope, operated on the island of La Palma by the ING in the Spanish Observatorio del Roque de los Muchachos of the Instituto de Astrofisica de Canarias.

\end{document}